**Full citation:** Felizardo, K.R., MacDonell, S.G., Mendes, E., & Maldonado, J.C. (2012) A systematic mapping on the use of visual data mining to support the conduct of systematic literature reviews, *Journal of Software* 7(2), pp.450-461.
doi:10.4304/jsw.7.2.450-461

# A Systematic Mapping on the use of Visual Data Mining to Support the Conduct of Systematic Literature Reviews

Katia R. Felizardo[1]

[1]*University of São Paulo/Department of Computer Science, São Carlos, Brazil*
*Email: katiarf@icmc.usp.br*

Stephen G. MacDonell[2] and Emília Mendes[3] and José Carlos Maldonado[1]

[2]*SERL, Comp. & Math. Sciences, AUT University, Auckland, New Zealand*
[3]*College of Information Technology, Zayed University, Dubai, United Arab Emirates*
*Email: stephen.macdonell@aut.ac.nz, emilia.mendes@zu.ac.ae, jcmaldon@icmc.usp.br*

## Abstract

*A systematic literature review (SLR) is a methodology used to find and aggregate all relevant existing evidence about a specific research question of interest. Important decisions need to be made at several points in the review process, relating to search of the literature, selection of relevant primary studies and use of methods of synthesis. Visualization can support tasks that involve large collections of data, such as the studies collected, evaluated and summarized in an SLR. The objective of this paper is to present the results of a systematic mapping study (SM) conducted to collect and evaluate evidence on the use of a specific visualization technique, visual data mining (VDM), to support the SLR process. We reviewed 20 papers and our results indicate a scarcity of research on the use of VDM to help with conducting SLRs in the software engineering domain. However, most of the studies (16 of the 20 studies included in our mapping) have been conducted in the field of medicine and they revealed that the activities of data extraction and data synthesis, related to conducting the review phase of an SLR process, have more VDM support than other activities. In contrast, according to our SM, previous studies using VDM techniques with SLRs have not employed such techniques during the SLR's planning and reporting phases.*

**Index Terms:** Systematic Mapping; Systematic Literature Review; Visual Data Mining

## 1. INTRODUCTION

A systematic literature review (SLR), or simply systematic review (SR), is used to provide a complete and fair evaluation of the state of all relevant research available on a specific topic of interest. An SLR process includes three main phases [1]: (i) planning; (ii) conducting the review; and (iii) reporting the review. During the first phase the need for a review is identified and the review protocol is developed. In the second phase the identification of a research question, the selection of primary studies complying with inclusion criteria, studies' quality assessment, data extraction and data synthesis tasks are executed. Finally, in the last phase, the results of the review are reported. A systematic mapping (SM), also known as a scoping review, is a more open form of SLR, providing an overview of a research area to assess the amount of existent evidence on a topic of interest [1].

SLRs were introduced in the software engineering (SE) field in 2004 [1] and have since gained increasing popularity among SE researchers [2]. However, in spite of its usefulness and growing importance, the SLR process remains challenging, time-consuming and manually conducted. Due to the necessarily comprehensive and rigorous nature of an SLR, exhaustive searches for relevant primary studies are required. However, recent studies have shown that these represent an especially difficult activity part of an SLR process [3, 4]. Moreover, the selection of primary studies is also arduous when faced with large volumes of candidate primary studies. Consequently, this leads to difficulties in reading, evaluating, and synthesizing the state of the art of a current topic of interest [3]. The adoption of approaches to aggregate research outcomes to provide a balanced and objective summary of research evidence is also a challenge [5].

The process of extracting patterns or non-trivial knowledge from structured data is known as data mining (DM) [6]. Information visualization uses one or more different techniques to create visual representations (images) to support the exploration of data sets [6]. Visual Data Mining (VDM), then, is the association of DM algorithms and information visualization techniques that supports visualization and interactive data exploration [7].

VDM techniques enable the use of human's strong visual processing abilities to support knowledge discovery [6] i.e., in the SLR context, to discover for example the relevant primary studies. Malheiros et al. [8] highlighted that VDM might significantly improve the SLR process. They found that the VDM approach accelerated the selection process and also improved the precision of the selection of relevant studies. One of our own previous studies has employed VDM in an Evidence-based Software Engineering (EBSE) context [9]. We proposed an approach based on VDM to support the categorization and classification stages of a systematic mapping (SM) process and found the application of VDM to be very relevant in the context of SM. Based on these previous findings we contend that VDM can help reviewers manage data during processes such as the classification and summarization of documents, and extraction of information. Our interest is in the identification of VDM techniques that could be used to support the activities involved in the SLR process, as well as the phases to which these techniques should be applied. Thus, considering the importance of knowing and understanding whether or not it is feasible or beneficial to use VDM techniques to support the SLR process, we propose a systematic mapping in order to analyze the use of VDM techniques to support the different stages and activities of the SLR process.

The remainder of the paper is organized as follows. Section 2 provides a brief overview of VDM. Section 3 describes our SM. A discussion of the results is given in Section 4, followed by conclusions and comments on future work in Section 5.

## 2. OVERVIEW OF VDM

VDM is an analysis approach that leverages the advantages of both DM and visualization [6]. A distinction is that VDM is a human guided process, whereas DM algorithms automatically analyze a textual data set searching for useful information [7]. Text mining (TM) refers to the application of DM to textual information or unstructured data, i.e., textual documents written in natural language. Finally, Visual Text Mining (VTM) is VDM applied to a collection of documents [10]. The exploration and analysis of large sets of documents (primary studies, for example) can be supported by VDM techniques [10, 6], which are briefly described as follows [11]:

- **Categorization:** Categorization (also referred to as classification) involves identifying the main themes of a document by considering the document's content in terms of pre-defined categories. Generally, text categorization uses supervised learning algorithms [12] to perform the classification automatically on unknown examples (unlabeled documents), based on the rules learnt from training documents or known examples (labeled documents).

- **Clustering:** Clustering groups similar documents but it differs from categorization that uses predefined categories. Latent Dirichlet Allocation (LDA) [13] is an example of a learning algorithm used for clustering documents, which also extracts topics (representative labels) from these documents.

- **Information retrieval (IR):** Information retrieval (IR) is a field of study that investigates mechanisms for searching information in text, images, video and other data. Document retrieval is a branch of IR where the information is stored primarily in the form of text. Document retrieval is the computerized process of locating documents based on the matching of a user's query request against a set of records – in this case unstructured texts. It produces a ranked list of documents that are relevant in response to the request. Boolean, probabilistic and natural language models are commonly used to support document retrieval [14].

- **Information extraction (IE):** IE is a technique used to detect specific pieces of information in text documents (unstructured format) and present it in a structured format. Generally, the information is detected using pattern matching, which identifies strings – words or phrases (also called patterns) related to pre-defined types of entity (e.g. names, time, place, events, among others) and their relationships within a text.

- **Summarization:** The goal of summarization is to reduce the length and detail of a document while retaining its main points and overall meaning. One of the strategies most widely used is sentence extraction, which extracts important concepts by statistically weighting the sentences.

## 3. SYSTEMATIC MAPPING

In general, a SM begins with a planning phase, which includes formulation of research questions and definition of inclusion and exclusion criteria, followed by search and screening of primary studies. The data extraction activity for a SM is broader than the data extraction process for an SLR and the analysis of a mapping does not include the use of in-depth analysis techniques, such as meta-analysis, but rather totals and summaries. Graphic representations also can be used to summarize the data [1, 15].

### A. Research Questions

An approach commonly used to formulate SM research question(s) is to use the PIOC criteria. Using PIOC the research questions are structured in four facets: (i) population; (ii) intervention; (iii) outcome; and (iv) context [16]. The PIOC attributes of the research questions utilized by us are shown in Table 1.

**TABLE I.** SUMMARY OF PIOC.

| Population | systematic literature reviews (SLR) |
|---|---|
| Intervention | visual data mining (VDM) techniques |
| Outcomes | VDM techniques used to support the SLR process and the SLR phases to which they have been applied |
| Context | with a focus on SLR |

Exploring and synthesizing vast volumes of data, such as the studies collected in an SLR, can be difficult tasks. Information visualization and VDM can help reviewers to deal with a large amount of information [6]. The aim of this paper is to understand of the use of VDM to support the SLR process, examining the types of techniques which may be applied in each activity. More specifically, the research question (RQ) that needs to be addressed by our SM is: *"What evidence is there of VDM techniques being applied to help in the SLR process?"*. Our RQ can be decomposed into the following sub-questions:

- **RQ 1.1:** In what phases of the SLR process has VDM been applied?

- **RQ 1.2:** What VDM techniques have been used to support the SLR process?

## B. Identification of Relevant Literature

The purpose of an SM is to conduct a review of relevant studies to assess the quantity of evidence existing to address the RQs [15]. The process needs to be rigorous and unbiased and it often involves a wide coverage of sources, such as online databases, journals and conferences. In order to minimize bias and to maximize the number of sources examined, a pre-defined strategy to identify potential primary studies is required. Ours is described as follows.

**TABLE II.** TERMS DERIVED FROM PIOC.

| Population | systematic literature review (SLR) |
|---|---|
| Interventions | visual data mining |
| Outcomes | VDM / systematic literature review/ SLR |
| Context | systematic literature review (SLR) |

**TABLE III.** TERMS DERIVED FROM KEYWORDS FOUND IN PAPERS.

| Author(s) | Year | Keywords |
|---|---|---|
| Keim [6] | 2002 | information visualization, visual data mining, VDM, visual data exploration |
| Oliveira & Levkowitz [7] | 2003 | information visualization, visual data exploration, visual data mining, survey |
| Lopes et al. [10] | 2007 | visual text mining, VTM, information visualization, text mining |
| Paulovich & Minghim [17] | 2008 | text visualization, document visualization, visual knowledge discovery |
| Kitchenham et al. [18] | 2009 | systematic literature review, systematic literature reviews |
| Kitchenham et al. [19] | 2010 | review of studies, structured review, systematic review, literature review, literature analysis, in-depth survey, literature survey, meta-analysis, past studies, empirical body of knowledge, overview of existing research, body of published research |

The strategy used to construct search terms was composed of 4 stages: 1. Identify the main terms considering the research questions (PIOC) – (see Table 2); 2. Identify synonyms or alternative words or abbreviations for major terms considering keywords found in papers on VDM or SLRs (see Table 3); 3. Use the Boolean OR to incorporate synonyms or alternative words or abbreviations (see Table 4) and; 4. Finally, use the Boolean AND to link the major terms (see Table 5).

To reduce the likelihood for bias, we validated our search string conducting a pilot search using a single database - IEEE Xplore, a search engine specialized in academic material, before applying it to all the selected databases. The pilot search intended to assess the completeness of the string, measured as the number of known relevant studies indexed by this database that were retrieved using that string. This search string was created using boolean logic to ensure comparability of results across databases. Once the pilot was carried out, we revisited the search string and new terms were not included.

After the definition of the search terms, the process of identifying the relevant literature was initiated. Our search was based on the following electronic databases: ACM Digital Library; IEEE Xplore; Web of Science; Scopus and Springer Link. We chose these databases based on existing literature on SLRs in SE. The authors decided not include the PubMed database (i.e., a freely accessible database on life sciences and biomedical topics – research area where the SLR was originally employed) because both Scopus and Web of Science already provided sufficient coverage of different disciplines, such as social sciences, technology and medicine.

Details of all potentially relevant primary studies were stored using the JabRef software, an open source bibliography reference manager. We used the "export" feature available in many electronic databases to export automatically the details of all potential primary studies (i.e., title, author(s), abstract, keywords, year of publication and the name of the data source) to JabRef. Information from databases that did not support exporting to JabRef was manually recorded by the first author.

**TABLE IV.** CONCATENATION OF ALTERNATIVE WORDS USING BOOLEAN OR.

| No. | Main topic | Results |
|---|---|---|
| 1 | visual data mining | (information visualization **OR** visual data mining **OR** VDM **OR** visual data exploration **OR** text mining **OR** document visualization **OR** visual text mining **OR** VTM **OR** visual knowledge discovery) |
| 2 | systematic literature review | (systematic literature review **OR** SLR **OR** systematic review **OR** systematic reviews **OR** literature review **OR** review of studies **OR** structured review **OR** literature analysis OR in-depth survey **OR** literature survey **OR** meta-analysis **OR** past studies **OR** empirical body of knowledge **OR** overview of existing research **OR** body of published research) |

**TABLE V.** CONCATENATION OF ALL POSSIBLE WORDS USING THE BOOLEAN AND.

| Final String |
|---|
| (information visualization OR  visual data mining OR VDM OR visual data exploration OR text visualization OR text mining OR document visualization OR visual text mining OR VTM OR visual knowledge discovery) **AND** (systematic literature review OR systematic literature reviews OR SLR OR  systematic review OR literature review OR review of studies OR structured review OR literature analysis OR in-depth survey OR literature survey OR meta-analysis OR past studies OR empirical body of knowledge OR overview of existing research OR body of published research) |

## C. Inclusion and Exclusion Criteria for Study Selection

Studies were **included** in the SM if they met the following inclusion criteria (IC):

- **(IC1)** The study investigated the use of VDM techniques to support the SLR process.

With regard to the exclusion criteria (EC), studies were excluded if:

- **(EC1)** The study investigated the use of VDM techniques, but did not consider their application in the SLR context and;

- **(EC2)** the study did not investigate the use of VDM techniques.

## D. Study Selection

The study selection activity was conducted in three phases. During phase 1, the first author applied the search strategy to identify potential primary studies. Duplicate papers were excluded. After this, in phase 2, two other researchers performed screening on the titles and abstracts for the purpose of deciding whether to include or exclude a study. The individual results were compared and the disagreement rate was 1.85%, i.e., only three studies did not have the same classification by both researchers. The full texts of these papers were read independently by the researchers and the disagreements were resolved by consensus between them. Therefore, in phase 3, the full text of the primary studies included in the preliminary selection was obtained. The first author read in detail the full text of each primary study included in the preliminary selection in order to decide whether to include or exclude the study. The same studies were analysed by one other researcher, who independently read the papers, and again disagreements were resolved by discussion and consensus (only one study did not have the same classification by both researchers). The primary studies included in the final selection correspond to the relevant papers that meet the RQ addressed by this SM.

The search for primary studies was conducted according to the guidelines described above. Initially, we sought potential primary studies in the databases. As a result, 304 studies were identified, including 112 duplicates. Next, we selected the primary studies by reading their titles and abstracts and applying the inclusion and exclusion criteria. As a result, a total of 30 studies were selected and 162 were excluded. Finally, the 30 papers were read in full and inclusion and exclusion criteria were again applied, resulting in 10 studies being rejected. Thus, we identified 20 relevant studies from the five sources that we searched (see Tables 6 and 7 and Figure 1). Note that the studies used previously to identify search keywords were not included in our SM because they did not address the use of VDM in the SLR process, rather, the topic addressed was VDM or SLR, exclusively.

TABLE VI.    SUMMARY OF RESEARCH RESULTS.

| Database Name | N. of Search Results |
|---|---|
| IEEE Xplore | 18 |
| ACM DL | 29 |
| Springer Link | 15 |
| Web of Science | 75 |
| Scopus | 167 |
| **Total** | **304** |
| **N. of duplicates found (phase 1)** | 112 |
| **N. of studies excluded (phase 2)** | 162 |
| **N. of studies excluded (phase 3)** | 10 |
| **N. of relevant studies** | 20 |

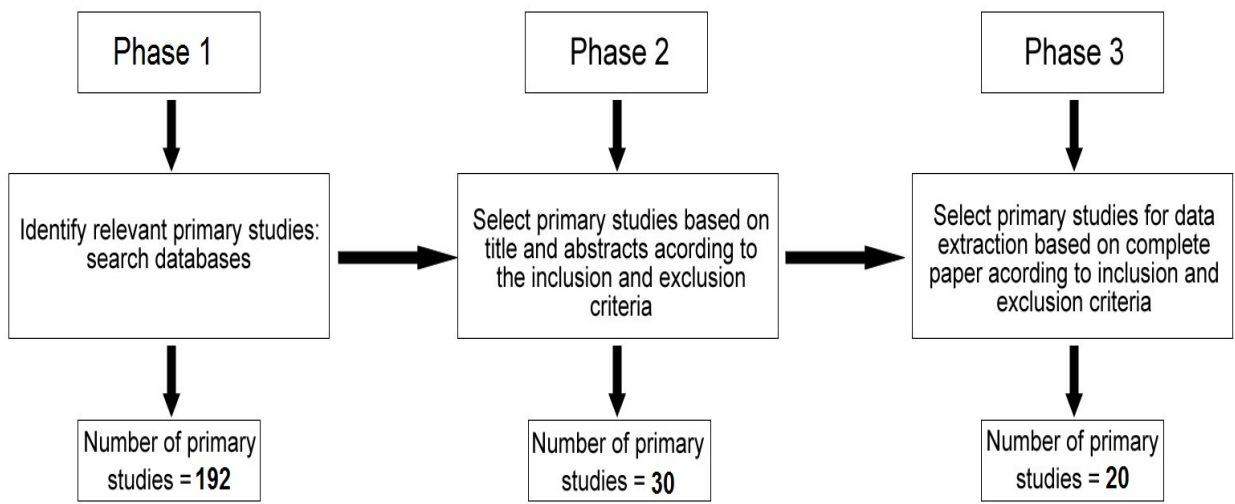

**Figure 1.** Phases of the study selection activity.

**TABLE VII.** LIST OF SELECTED STUDIES.

| Study ID | Title | Authors | Year | Country of Institution | Study Ref. |
|---|---|---|---|---|---|
| S1 | A Visual Text Mining approach for Systematic Reviews | V. Malheiros and E. Hohn and R. Pinho and M. Mendonca | 2007 | Brazil | [8] |
| S2 | Co-clustering approaches to integrate lexical and bibliographical information | F. Janssens and P. Glenisson and W. Glanzel and B. De Moor | 2005 | Belgium | [20] |
| S3 | Combining mining and visualization tools to discover the geographic structure of a domain | J. Mothe and C. Chrisment and T. Dkaki and B. Dousset and S. Karouach | 2006 | France | [21] |
| S4 | Conceptual biology, hypothesis discovery, and text mining: Swanson's legacy | T. Bekhuis | 2006 | USA | [22] |
| S5 | Data mining of cancer vaccine trials: A bird's-eye view | X. Cao and K. B. Maloney and V. Brusic | 2008 | USA | [23] |
| S6 | Data mining techniques to enable large-scale exploratory analysis of heterogeneous scientific data | P. Chopra and D. L. Bitzer and S. Heber | 2009 | USA | [24] |
| S7 | Enhancing the Literature Review Using Author-Topic Profiling | A. Kongthon and C. Haruechaiyasak and S. Thaiprayoon | 2008 | Thailand | [25] |
| S8 | Exploitation of ontological resources for scientific literature analysis: searching genes and related diseases | A. Jimeno-Yepes and R. Berlanga-Llavori and D. Rebholz-Schuhmann | 2009 | UK | [26] |
| S9 | Extracting information from textual documents in the electronic health record: a review of recent research | S. M. Meystre and G. K. Savova and K. C. Kipper-Schuler and J. F. Hurdle | 2008 | USA | [27] |
| S10 | Extracting knowledge from genomic experiments by incorporating the biomedical literature | J. P. Sluka | 2002 | USA | [28] |
| S11 | Finding relevant PDF medical journal articles by the content of their figures | A. Christiansen and D. -J. Lee and Y. Chang | 2007 | USA | [29] |
| S12 | Identification of differentially expressed proteins using automatic meta-analysis of proteomics-related articles | E. A. Ponomarenko and A. V. Lisitsa and I. Petrak and S. A. Moshkovskii and A. I. Archakov | 2009 | Russia | [30] |
| S13 | Parameterized Contrast in Second Order Soft Co-occurrences: A Novel Text Representation Technique in Text Mining and Knowledge Extraction | A. H. Razavi and S. Matwin and D. Inkpen and A. Kouznetsov | 2009 | Canada | [31] |
| S14 | Reconstruction of protein-protein interaction pathways by mining subject-verb-objects intermediates | M. H. T. Ling and C. Lefevre and K. R. Nicholas and F. Lin | 2007 | Australia | [32] |
| S15 | Research profiling: Improving the literature review | A. L. Porter and A. Kongthon and J.-C. Lu | 2002 | USA | [33] |
| S16 | Seeding the survey and analysis of research literature with text mining | D. Delen and M. D. Crossland | 2008 | USA | [34] |
| S17 | SENT: Semantic features in text | M. Vazquez and P. Carmona-Saez and R. Nogales-Cadenas and M. Chagoyen and F. Tirado and J. M. Carazo and A. Pascual-Montano | 2009 | Spain | [35] |
| S18 | Supporting Systematic Reviews Using Text Mining | S. Ananiadou and B. Rea and N. Okazaki and R. Procter and J. Thomas | 2009 | UK | [36] |
| S19 | SYRIAC: The systematic review information automated collection system a data warehouse for facilitating automated biomedical text classification | J. J. Yang and A. M. Cohen and M. S. McDonagh | 2008 | USA | [37] |
| S20 | The next generation of literature analysis: Integration of genomic analysis into text mining | M. Scherf and A. Epple and T. Werner | 2005 | Germany | [38] |

## E. Study Quality Assessment

In order to analyse the quality of the included primary studies we developed a checklist containing 5 questions (see Table 8). Our checklist was adapted from the generic quality criteria created by Kitchenham [1]. For each question in the checklist, the following scale-point was applied: Yes – 1 point; No – 0 points; Partially – 0.5 point. The total quality score therefore fell into the range between: 0–1.0 (very poor); 1.1 – 2.0 (fair); 2.1 – 3.0 (good); 3.1 – 4.0 (very good) and 4.1 – 5.0 (excellent). We also used the quality score as a basis for decisions on the inclusion of primary studies. Since there is no agreed definition of study "quality" [1], based on the quality procedures decided by our mapping team, the cutoff scale-point selected by us to exclude studies from the list of selected studies was 2.0 (very poor or fair quality). Thus, we extracted data from all 20 studies selected previously which as they had all been classified as good, very good or excellent quality. Table 9 reports the evaluation of each study against each item of the checklist and gives a summary of the scores.

**TABLE VIII.** STUDY QUALITY CHECKLLIST.

| N. | Item | Answer |
|---|---|---|
| 1 | Is it clear which visual mining technique was used? | Yes/No |
| 2 | Is the visual mining technique fully defined (visual representation, exploration strategy and tool associated)? | Yes/No/Partially |
| 3 | Is it clear to which phase(s) of the SLR process the visual mining technique was applied? | Yes/No/Partially |
| 4 | Is it clear how the visual mining technique was used? | Yes/No/Partially |
| 5 | Has the use of the visual mining technique been validated? | Yes/No/Partially |

**TABLE IX.    DETAILS OF THE QUALITY SCORE.**

| Study ID | Q1 | Q2 | Q3 | Q4 | Q5 | Total Score | Study ID | Q1 | Q2 | Q3 | Q4 | Q5 | Total Score |
|---|---|---|---|---|---|---|---|---|---|---|---|---|---|
| S1 | Y | Y | Y | Y | Y | 5.0 | S11 | Y | Y | Y | Y | Y | 5.0 |
| S2 | Y | Y | N | Y | Y | 4.0 | S12 | Y | Y | N | P | Y | 3.5 |
| S3 | Y | Y | N | Y | Y | 4.0 | S13 | Y | Y | Y | Y | Y | 5.0 |
| S4 | P | Y | N | Y | Y | 3.5 | S14 | Y | Y | N | Y | Y | 4.0 |
| S5 | P | Y | N | Y | Y | 3.5 | S15 | Y | Y | P | Y | Y | 4.5 |
| S6 | Y | Y | Y | Y | Y | 5.0 | S16 | Y | Y | N | Y | Y | 4.0 |
| S7 | Y | Y | N | Y | Y | 4.0 | S17 | Y | Y | P | Y | Y | 4.5 |
| S8 | Y | P | N | P | P | 2.5 | S18 | Y | Y | Y | Y | Y | 5.0 |
| S9 | Y | P | N | P | P | 2.5 | S19 | Y | Y | Y | Y | Y | 5.0 |
| S10 | Y | Y | Y | Y | Y | 5.0 | S20 | Y | P | P | P | N | 2.5 |

## F. Data Extraction

With the final set of primary studies decided upon and their quality assessed, the data extraction activity was carried out on all 20 papers (as they had all passed the screening process). The first author was responsible for extracting the data and completing the associated forms, the content of which is summarized in Table 10. For validation purposes, a sample comprising 20% of the total number of primary studies was selected randomly and had their data extracted by the one other researcher. There was a high level of agreement between the first author and the other researcher (95%), with the difference being discussed until consensus was reached. The disagreement related to the classification of the LDA technique (primary study S7), which was classified at first by the first author, as a new category. Considering that LDA is an algorithm used for clustering documents, it was inserted in the "clustering" category after discussion.

## G. Data Synthesis and Results of the Systematic Mapping

The data synthesis activity involves compiling the data extracted from each primary study included in the SM. Our results were summarized to present an overview of the findings, thus, our study is a scoping study [1] that maps out the VDM techniques that have been used to support the SLR process. We planned to perform the data synthesis for our RQs using tables (totals and summaries) and visual representations (graphs). The graphs were chosen because they are an alternative visual representation that can be used to represent findings, showing connections between concepts and findings (e.g. VDM techniques x SLR phases/activities). One example of the use of graphs to show connections among the primary studies' findings in an SLR can be found in [39].

**TABLE X.    SUMMARY OF EVIDENCE FOR VISUAL DATA MINING TECHNIQUES.**

| Study ID | VDM Techniques | SLR Support | Area |
|---|---|---|---|
| S1 | This paper suggests the use of document map (visual representation) to support the selection activity. The authors created 3 VDM strategies: clusters; clusters with label (representative topics about the documents grouped in the cluster) and to change the colour of the documents in the map to represent the occurrence of specific terms. | B2 | Software Engineering |
| S2 | The study explores how a methodology of indexing full-text scientific articles combined with an exploratory statistical analysis can improve on bibliometric approaches to mapping science. | B5 | Social Sciences |
| S3 | This article presents the Tétralogie platform, which allows a user to interactively discover trends in scientific research and communities from large textual collections that include information about geographical location (information extraction and document categorization). The results are displayed through spreadsheets, histograms, graphs, 4D-views; geographic maps and networks. | B4; B5 | Medicine |
| S4 | It shows a search strategy to find "undiscovered public knowledge". In this strategy, the knowledge discovery of different domains is crossed in a single system in order to establish an unexpected link between two terms. | B1 | Medicine |
| S5 | It describes a data mining approach that enables rapid extraction of complex data from the major clinical trial repository (extraction, summarization and visualization of extracted knowledge from cancer vaccine clinical trials data). The information is presented using graph format (bars, scatter plot). | B4; B5 | Medicine |
| S6 | This PhD work addressed three aspects of data mining in biological datasets: clustering, categorization and meta-analysis of microarray (i.e., chips that are used to detect the RNA levels of genes). | B4; B5 | Medicine |
| S7 | It applies Latent Dirichlet Allocation (LDA) to generate topics to discover author-topic relations from text collections. The results are presented in tables. | B4 | Medicine |
| S8 | The paper is an overview about the use of Text Mining solutions for information retrieval (IR) and information extraction (IE). IR: query reformulation; query expansion (it uses ontological resources); query refinement; IE: structured output expressed by a template. | B1; B4 | Medicine |
| S9 | This paper reviews the advances of information extraction from Electronic Health Record (EHR) documents. Examples of extraction: extraction of codes (International Classification of Diseases - ICD); extraction of information for decision-support and enrichment of the EHR, information extraction for surveillance, automated terminology management and de-identification of clinical text (i.e., hidden or replace PHI – Protected Health Information). | B4 | Medicine |
| S10 | This paper presents the PDQ_MED (Pair-wise Data Query to MEDLINE), a program based on the assumption that if two genes are found to be related under an experimental paradigm, such as a gene chip experiment, then any literature which relates the two genes is of interest. | B1 | Medicine |

| Study ID | VDM Techniques | SLR Support | Area |
|---|---|---|---|
| S11 | This PhD work describes a search strategy (document retrieval) that uses content-based image retrieval (CBIR) techniques to search for relevant documents using the content of figures in a document instead of keyword search. | B1 | Medicine |
| S12 | It proposes an algorithm for automatic meta-analysis of proteomic publications based on evaluation of the frequency of protein names found in text sets. | B5 | Medicine |
| S13 | A statistical representation method based on second order co-occurrence vectors for knowledge extraction. It can be used to classify medical abstracts for systematic reviews in two classes: (i) positive class (relevant articles) and (ii) negative class (not relevant articles). | B2 | Medicine |
| S14 | MontyLingua is an information extraction tool that uses a generalization-specialization paradigm, where text was generically processed to a suitable intermediate format before domain-specific data extraction techniques are applied at the specialization layer. | B4 | Medicine |
| S15 | Using the "research profiling" topical relationships, research trends and active organizations and individuals whose research relates to one's own interests can be discovered. | B4; B5 | Systems Engineering /Medicine |
| S16 | It applies text categorization to organize abstracts into logical categories or topic clusters. | B4 | Information Systems |
| S17 | The SENT (Semantic Features in Text) tool uses non-negative matrix factorization to identify topics in the scientific articles related to a collection of genes or their products, and use them to group and summarize these genes. | B5 | Medicine |
| S18 | Searching can be improved by using query expansion (ontologies and thesauri could be used). Screening can be improved by using document clustering which groups documents into topics. Synthesizing can be improved by using an adaptable multi-document summarization driven by user defined viewpoints. | B1; B2; B5 | Social Sciences |
| S19 | The SYRIAC (SYstematic Review Information Automated Collection System) creates training datasets for SLR text mining research (automatic document classification algorithms). | B2 | Medicine |
| S20 | It shows an overview about the challenges of identification, description and classification of relations between biological entities (genes, proteins or diseases) from free text. | B4 | Medicine |

**SLR Support**

A. The stages associated with ***planning the review*** are:

A1. Identification of the need for a review

A2. Development of a review protocol

B. The stages associated with ***conducting the review*** are:

B1. Identification of research

B2. Selection of primary studies

B3. Study quality assessment

B4. Data extraction and B5. Data synthesis

*C. **Reporting the review** is a single stage phase*

**TABLE XI.**   PHASES OF SLR PROCESS SUPPORTED BY VISUAL DATA MINING TECHNIQUES (TABLE REPRESENTATION).

| SLR Support | | Studies ID |
|---|---|---|
| Planning the Review (A) | **A1** | - |
| | **A2** | - |
| Conducting the Review (B) | **B1** | S4; S8; S10; S11; S18 (5) |
| | **B2** | S1; S13; S18; S19 (4) |
| | **B3** | - |
| | **B4** | S3; S5; S6; S7; S8; S9; S14; S15; S16; S20 (10) |
| | **B5** | S2; S3; S5; S6; S12; S15; S17; S18 (8) |
| Reporting the Review (C) | **C** | - |

**SLR Support**

A. The stages associated with ***planning the review*** are:

   A1: Identification of the need for a review

   A2: Development of a review protocol

B. The stages associated with ***conducting the review*** are:

   B1: Identification of research

   B2: Selection of primary studies

   B3: Study quality assessment

   B4: Data extraction

   B5: Data synthesis

C: ***Reporting the review*** is a single stage phase

Moreover, graphic representations of results are often easier for readers to understand than tables [40] and so may be an effective reporting mechanism [1]. A graph is an abstract data structure that consists mainly of a finite set of ordered pairs, edges and nodes. The nodes represent objects (that can be tangible or intangible depending on the application) and they are connected by edges that can refer to some common shared aspect [41]. The information contained in the SM tables was restructured to be used in an open source tool called PEx-Graph [42], which creates the graphs containing the same information presented in the tables.

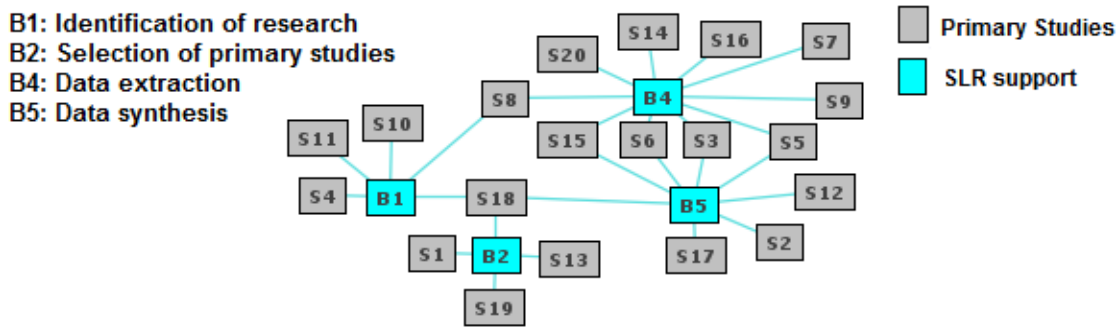

**Figure 2.** Phases of SLR process supported by Visual Data Mining techniques (graph representation) [1].

In the case of **RQ 1.1** (i.e., In what phases of the SLR process has VDM been applied?), synthesized data from all 20 studies (see Table 11 and Figure 2) show that activities B1, B2, B4 and B5 of the *conducting the review* phase (i.e., identification of research; selection of primary studies; data extraction; and data synthesis, respectively) have received VDM support. The collected evidence indicates that the activities that have more support are B4, data extraction (S3; S5; S6; S7; S8; S9; S14; S15; S16; S20 – 10 studies: 50%) and B5, data synthesis (S2; S3; S5; S6; S12; S15; S17; S18 – 8 studies: 40%).

The results also indicate that activities B1, identification of research (S4; S8; S10; S11; S18 – 5 studies: 25%) and B2, selection of primary studies (S1; S13; S18; S19 – 4 studies: 20%) have also received a degree of VDM support. It is interesting to note that, in contrast, there is no evidence of VDM support for the *planning the review* phase, its respective activities A1 and A2 (i.e., identification of the need for a review, development of a review), or the *reporting the review* phase. Most of the studies (16 of the 20, or 80%) have been conducted in the field of medicine (S3; S4; S5; S6; S7; S8; S9; S10; S11; S12; S13; S14; S15; S17; S19; S20). In the SE domain we found only one study (S1), which supports the selection of primary studies activity. This finding suggests that the remaining phases and activities of the SLR process in SE are generally conducted manually. A possible explanation for this might be that medical research has utilized the evidence-based paradigm for the last two decades and in that field of research the SLR is recognized as one of the key components of the Evidence-Based Medicine (EBM) paradigm. In contrast, SLRs were introduced in SE in 2004, as a method for conducting secondary studies as part of the emerging EBSE paradigm.

**TABLE XII.** SUMMARY OF VISUAL DATA MINING TECHNIQUES AND THEIR RESPECTIVE STUDIES (TABLE REPRESENTATION).

| VDM Technique | SLR Process Stage |
|---|---|
| Clustering | B2 (S1; S18);  B4(S6; S7) |
| Document Categorization / Classification | B2 (S13; S19); B4 (S6; S16; S20) |
| Document Map | B2 (S1) |
| Document Retrieval (search strategy) | B1 (S4; S8; S10; S11; S18) |
| Information Extraction | B4 (S3; S5; S8; S9; S14; S15) |
| Document Summarization | B5 (S2; S3; S5; S6; S12; S15; S17; S18) |

**TABLE XIII.** SUMMARY OF VISUAL DATA MINING TECHNIQUES USED TO SUPPORT SLR PROCESS (TABLE REPRESENTATION).

| VDM Technique | SLR Process Stage |
|---|---|
| Document Retrieval (S4; S8; S10; S11; S18 – 5 studies) | B1 – Identification of research (5 studies) |
| Clustering (S1; S18 – 2 studies)<br>Document Categorization / Classification (S13; S19 – 2 studies)<br>Document Map (S1 – 1 study) | B2 –  Selection of primary studies (5 studies) |
| Clustering (S6; S7 – 2 studies)<br>Document Categorization / Classification  (S6; S16; S20 – 3 studies)<br>Information Extraction (S3; S5; S8; S9; S14; S15 – 6 studies) | B4 –  Data extraction (10 different studies) |
| Document Summarization (S2; S3; S5; S6; S12; S15; S17; S18 – 8 studies) | B5 –  Data synthesis (8 studies) |

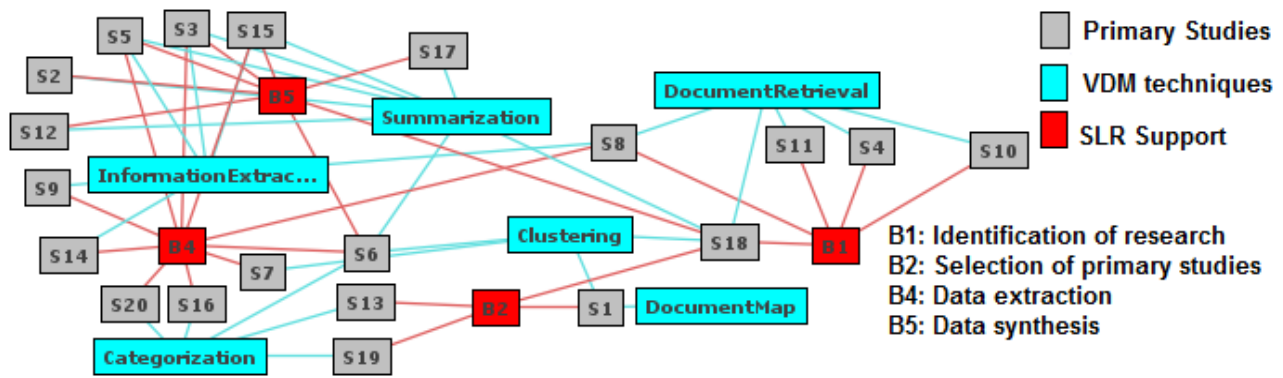

**Figure 3.** Summary of Visual Data Mining techniques used to support SLR process (graph representation).

In the case of **RQ 1.2** (i.e., What VDM techniques have been used to support the SLR process?), the collected evidence (Tables 12 and 13 and Figure 3) shows the use of various types of VDM techniques to support the same phase/activity. For example, activity B2, selection of primary studies, may be supported by clustering (S1; S18 – 2 studies: 10%); document map (S1 – 1 study: 5%); and document categorization/classification (S13; S19 – 2 studies: 10%) techniques. In the case of activity B4, data extraction, it has been supported by clustering (S6; S7– 2 studies: 10%); document categorization/classification (S6; S16; S20 – 3 studies: 15%); and information extraction (S3; S5; S8; S9; S14; S15 – 6 studies: 30%) techniques. The current study also found that VDM techniques such as clustering and document categorization/classification can be used to support different activities (i.e., B2, selection of the primary studies and B4, data extraction). Activity B1, identification of research, is supported by document retrieval techniques (S4; S8; S10; S11; S18 – 5 studies: 25%), and activity B5, data synthesis, exclusively by document summarization techniques (S2; S3; S5; S6; S12; S15; S17; S18 – 8 studies: 40%).

We did not apply a date range limit in our search, thus the results of the data synthesis summarized in Table 7 suggest that research on this issue began in 2002 (2 studies), and while no papers were published in 2003 and 2004, research has been steady from 2005. We can observe an increase in the number of primary studies related to VDM to support the SLR process since 2008. This seems to indicate an increasing interest in this topic of research. In terms of country of origin, the USA has contributed the majority of the studies, nine (45%). The number of papers retrieved for each area is as follows: information systems – 1 paper (5%); medicine – 16 papers (80%); social sciences – 2 papers (10%); SE – 1 paper (5%); and systems engineering – 1 paper (5%) (see Table 14 and Figure 4, where show the inter-relatedness of the studies is shown).

Some VDM tools that can be used to support activities part of an SLR process are: Projection Explorer – PEx (http://infoserver.lcad.icmc.usp.br/infovis2/PEx) [8]; Tétralogie platform [21]; Spotfire-DXP (http://www.tibco.com) [23]; PDQ_MED [28]; Muscorian, a biological text mining system [32];

Semantic Features in Text – SENT [35]; Assert (http://www.nactem.ac.uk/assert) [36] and SYstematic Review Information Automated Collection System – SYRIAC [37]. A comprehensive analysis of these tools is given in the references abovementioned. One of these tools - PEx [8], is specific to be used in the SE area. A summary of tools that extract information from textual documents (e.g., UMLS-based spelling error correction, REgenstrief eXtraction; CliniViewer; NLM's MetaMap and MedLEE, among others) is given in [27].

**TABLE XIV.** DISTRIBUTION OF PRIMARY STUDIES BY AREA (TABLE REPRESENTATION).

| Topic | Number of Papers |
|---|---|
| Information Systems | 1 |
| Medicine | 16 |
| Social Sciences | 2 |
| SE | 1 |
| Systems Engineering | 1 |

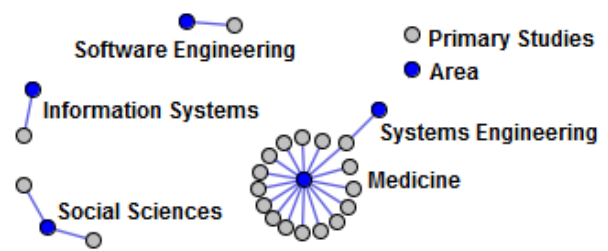

**Figure 4.** Distribution of primary studies by area (graph representation).

## 4. DISCUSSION OF THE RESULTS

There are phases of the SLR process (i.e., *planning* and *reporting the review*) that have not been addressed using VDM and more research on these topics needs to be undertaken. However, overall, the results are encouraging in the sense that there is good and growing evidence for the use of VDM to support the SLR process in medicine. It may be beneficial for SE researchers to investigate the advantages and limitations of VDM techniques as adopted in medicine to assess their potential value to those in the SE research community conducting SLRs.

The results of the SLR suggest that there is a predominance of document summarization techniques to support data synthesis and document retrieval techniques to support the identification of research. We found that only 1 of the 20 primary studies provided evidence of the use of VDM in the SE domain, which indicates a scarcity of research in this area and signals an opportunity for future research. Based on the evidence of our SM, the investigation of VDM to support the SLR process is an important issue for future research in SE. For example, the support provided by SE on-line indexing databases is not adequate because they do not support complex Boolean searches [5], but in this context document retrieval techniques may be investigated to minimize this problem.

There are no obvious choices to support the selection of primary studies and data extraction. The SM identified that the document classification technique employed to support the selection of primary studies may be equally applied to support data extraction. One possible explanation for this is that these activities are closely related in terms of assigning each primary study to one or more categories (i.e., included/excluded, or case study/controlled experiment/survey).

## 5. CONCLUSIONS

We have presented the results of an SM aimed at investigating the use of VDM to support the SLR process. This mapping has helped us understand the current state of research in VDM techniques to support those undertaking SLRs, and also in identifying research gaps and future directions. Based on our mapping it is clear that one of the research gaps lies in proposing specific VDM techniques to support SLRs in SE. Further work is needed to develop new VDM techniques and tools to support and automate specifically the different phases/activities of the SLR process in the SE domain. Although there are a number of issues where SE research differs from medical research (e.g. the SE domain has relatively little empirical research [15] it is our belief that we can learn from the experience in conducting SLRs in medicine to adapt/utilize the VDM techniques adopted in that domain (and summarized in this paper) for use in SE.

There is also an opportunity to extend Malheiros et al.'s work [8] which makes use of VDM techniques to support study selection in the SLR process as applied in the SE domain. New visualization techniques could be proposed and used together with the document map previously suggested by the authors. Moreover, it is important to investigate whether or not VDM plays a role in differentiating the performance of the selection activity. In other words, it is necessary to compare the performance of reviewers in selecting primary studies using the manual approach (i.e., reading abstracts/full papers) and using VDM techniques.

With respect to the threats to the validity of our SM, there are two main concerns: (i) we did not consider the alternative spelling "visualisation" in the construction of our search string. We believe that all relevant primary studies were identified, given the high percentage of duplicated primary studies found (36.8%). However, we cannot rule out that our choice (i.e. only the word "visualization" was used in our search string) could have led to some studies being missed; and (ii) we based our searches on a range of electronic databases, but other sources (e.g. reference lists from relevant primary studies; grey literature - technical reports, work in progress) were not searched.

## ACKNOWLEDGMENT

This research is supported by Brazilian funding agencies: CNPq (Processes n. 141972/2008-4; 201622/2009-2) and CAPES.

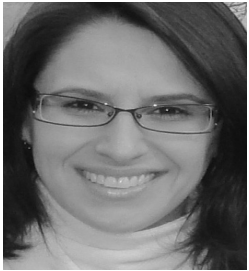

**Katia Romero Felizardo** received the M.Sc. degree in Computer Science from Federal University of São Carlos, Brazil, in 2003. Currently, she is a PhD student at the Computer Science Department of the Instituto de Ciências Matemáticas e de Computação, at the University of São Paulo, Brazil. Her research interests are in systematic literature review, information visualization and visual data mining.

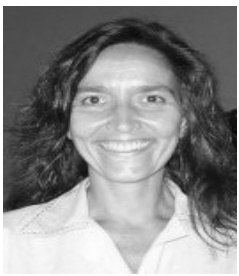

**Stephen G. MacDonell** received the PhD degree in software engineering from the University of Cambridge, Cambridge, U.K., in 1993. He is currently a Professor of Software Engineering and the Director of the Software Engineering Research Laboratory (SERL) at Auckland University of Technology (AUT), Auckland, New Zealand. He is currently engaged in research in software metrics and measurement, project planning, estimation and management, software forensics, and the application of empirical analysis methods to software engineering datasets. Prof. MacDonell is a member of the IEEE Computer Society, the Association for Computing Machinery (ACM), and the Editorial Board of Information and Software Technology.

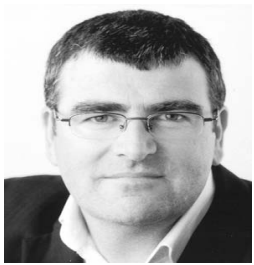

**Emilia Mendes** received the PhD degree in computer science from the University of Southampton, United Kingdom, in 1999 after working in the software industry for 10 years. She is an associate professor in computer science at the University of Auckland, New Zealand. She has active research interests in the areas of empirical Web and software engineering, evidence based research, hypermedia, computer science, and software engineering education, in which areas she has published widely, with over 100 refereed publications, including two books, one edited (2005) and one authored (2007). She is a member of the editorial boards of the International Journal of Web Engineering and Technology, the Journal of Web Engineering, the Journal of Software Measurement, the International Journal of Software Engineering and Its Applications, the Empirical Software Engineering Journal, and the Advances in Software Engineering Journal.

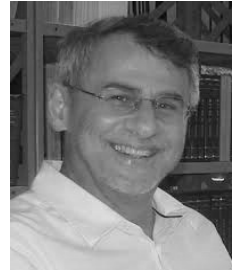

**José C. Maldonado** received the PhD degree in electrical engineering/ automation and control in 1991 from the University of Campinas (UNICAMP), Brazil. He worked at INPE from 1979 up to 1985 as a researcher. In 1985, he joined the Computer Science Department of the Institute of Mathematics and Computer at the University of São Paulo (ICMC-USP) where he is currently a faculty member. He has been a visiting scholar at the Technical University of Denmark, and at Purdue University, supported by the Fulbright Program and the Brazilian funding agencies CAPES and CNPq. His research interests are in the areas of software quality, software testing, and reliability. He has published more than 50 research papers and written one book. He is a member of the Brazilian Computer Society (SBC). He is a member of the IEEE Computer Society.